\documentclass[preprintnumbers,amsmath,amssymbm,prd]{revtex4}
\usepackage{epsfig}
\usepackage{graphicx}

\begin{document}
\title{Non-equatorial scalar rings supported by rapidly spinning Gauss-Bonnet black holes}
\author{Shahar Hod}
\affiliation{The Ruppin Academic Center, Emeq Hefer 40250, Israel}
\affiliation{ }
\affiliation{The Hadassah Institute, Jerusalem 91010, Israel}
\date{\today}

\begin{abstract}
\ \ \ Black-hole spacetimes that possess stationary equatorial matter rings are known to exist in general relativity. 
We here reveal the existence of black-hole spacetimes that support {\it non}-equatorial matter rings. 
In particular, it is proved that rapidly-rotating Kerr black holes in the dimensionless 
large-spin regime ${\bar a}>{\bar a}_{\text{crit}}=
\sqrt{\big\{{{7+\sqrt{7}\cos\big[{1\over3}\arctan\big(3\sqrt{3}\big)\big]-
\sqrt{21}\sin\big[{1\over3}\arctan\big(3\sqrt{3}\big)\big]}\big\}/12}}\simeq0.78$ can 
support a pair of non-equatorial massive scalar rings which are negatively coupled 
to the Gauss-Bonnet curvature invariant of the spinning spacetime 
(here ${\bar a}\equiv J/M^2$ is the dimensionless angular momentum of the central supporting 
black hole). We explicitly prove that these non-equatorial scalar rings are characterized by the dimensionless 
functional relation $-{{57+28\sqrt{21}\cos\big[{1\over3}\arctan\big({{1}\over{3\sqrt{3}}}\big)\big]}
\over{8(1+\sqrt{1-{\bar a}^2})^6}}
\cdot{{\bar\eta}\over{{\bar\mu}^2}}\to 1^{+}$ in the large-mass ${\bar\mu}\equiv M\mu\gg1$ regime (here 
$\{{\bar\eta}<0,\mu\}$ are respectively the non-trivial coupling parameter of 
the composed Einstein-Gauss-Bonnet-massive-scalar field theory and the proper mass of the supported non-minimally 
coupled scalar field). 
\end{abstract}
\bigskip
\maketitle

\section{Introduction}

The physically important no-hair conjecture \cite{NHC,JDB} for black holes has motivated the mathematical 
development of elegant non-existence theorems that established the intriguing fact 
that bound-state static scalar field configurations with minimal \cite{Bek1,Sot2,Her1,Sot3} (as well as with non-minimal \cite{BekMay,Hod1}) coupling to the spatially-dependent Ricci curvature scalar $R$ cannot be supported in the external regions of 
asymptotically flat black-hole spacetimes.

However, recent studies \cite{Sot5,Sot1,GB1,GB2} have explicitly demonstrated that the no-hair conjecture can be violated by asymptotically flat non-vacuum black holes with spatially regular horizons 
in generalized Einstein-Gauss-Bonnet-scalar field theories that contain a non-trivial 
interaction term $f(\phi){\cal G}$ between the scalar field $\phi$ of the theory 
and the Gauss-Bonnet invariant ${\cal G}$ of the curved spacetime 
\cite{Notechar,Hersc1,Hersc2,Hodsc1,Hodsc2,Moh,Hodnrx,HodKN}. 

The physically intriguing observation \cite{Sot5,Sot1,GB1,GB2} (see also \cite{ChunHer,SotN,Hodsg1,Hodsg2,Done,Herrecnum,BeCo}) 
that black holes in composed Einstein-Gauss-Bonnet-scalar field theories can be spontaneously scalarized is a direct consequence 
of the fact that the generalized Klein-Gordon differential equation that determines the spatio-temporal functional behavior of the 
non-minimally coupled scalar fields in the supporting curved black-hole spacetime 
contains an effective mass term of the form $-\bar\eta G$ [here $\bar\eta$ is the non-trivial Gauss-Bonnet-scalar-field dimensionless 
coupling parameter of the theory, see Eq. (\ref{Eq8}) below] which, 
depending on the physical parameters of the central supporting black hole, may become negative (thus representing an 
attractive black-hole-field interaction potential) in the highly curved near-horizon region. 

Interestingly, it has been demonstrated numerically \cite{ChunHer,SotN} and analytically \cite{Hodnccc} 
that composed Einstein-Gauss-Bonnet-scalar field theories are characterized by 
an {\it existence-line} which marks the sharp 
boundary between the familiar (scalarless) Kerr black-hole solutions of the Einstein field equations and 
composed spinning-black-hole-nonminimally-coupled-scalar-field hairy
configurations. 
In particular, it has been demonstrated \cite{ChunHer,SotN,Hodnccc} that the existence-line of the black-hole-field system describes 
the spin-dependent unique values $\bar\eta=\bar\eta({\bar a})$ of the non-trivial Gauss-Bonnet-scalar-field coupling parameter 
that allow spinning Kerr black holes to support spatially regular bound-state linearized configurations (linearized `scalar clouds' \cite{Hodlit,Herlit}) of the non-minimally coupled matter fields 
(here ${\bar a}\equiv J/M^2$ is the dimensionless spin parameter of the supporting black hole). 

As nicely discussed in \cite{Sot5,Sot1,GB1,GB2,ChunHer,SotN,Hodsg1,Hodsg2,Done,Herrecnum,BeCo}, one finds that the 
critical existence-line of the black-hole-field system is universal in the sense that different 
Einstein-Gauss-Bonnet-scalar field theories, which are characterized by different non-linear 
scalar coupling functions $\{f(\phi)\}$ that share the same weak-field functional expansion 
$f(\phi)=1+\bar\eta\phi^2/2+O(\phi^4)$, are characterized by the same critical boundary (critical existence-line) 
$\bar\eta=\bar\eta({\bar a})$ that separates bald (scalarless) Kerr black-hole spacetimes 
from non-trivial spinning-black-hole-scalar-field hairy spacetime configurations.

Intriguingly, it has recently been proved \cite{SotN} that, in Einstein-Gauss-Bonnet-scalar field theories in which the 
scalar field is {\it negatively} coupled ($\bar\eta<0$) to the Gauss-Bonnet curvature invariant of the spacetime, the 
spontaneous scalarization phenomenon of spinning Kerr black holes is restricted to the super-critical 
regime \cite{SotN,Hodca,Notelll}
\begin{equation}\label{Eq1}
{\bar a}>{1\over 2}\
\end{equation}
of the black-hole dimensionless spin parameter. 
In addition, it has been revealed that, in the $\bar\eta<0$ regime of the non-minimal Gauss-Bonnet-scalar-field 
coupling parameter, 
spontaneously scalarized spinning black holes are characterized 
by the near-critical {\it large}-coupling behavior \cite{SotN,Hodnccc}
\begin{equation}\label{Eq2}
-\bar\eta\to \infty\ \ \ \ \text{for}\ \ \ \ {\bar a}\to\Big({1\over 2}\Big)^+_{\text{crit}}\  .
\end{equation}

The main goal of the present paper is to explore, using {\it analytical} techniques, 
the physical and mathematical properties of composed 
spinning-Kerr-black-hole-massive-scalar-field cloudy configurations in the negative large-coupling regime. 
In particular, below we shall explicitly prove that the addition of proper masses to the non-trivially 
coupled scalar fields allows rapidly-rotating Kerr black holes [see Eq. (\ref{Eq18}) below] to 
support a pair of {\it non}-equatorial scalar rings which are negatively coupled 
to the Gauss-Bonnet invariant of the curved spacetime. 
Interestingly, we shall prove that the composed Kerr-black-hole-non-equatorial-scalar-rings cloudy configurations 
are characterized by the double {\it large}-coupling-{\it large}-mass behavior 
\begin{equation}\label{Eq3}
-\bar\eta\to \infty\ \ \ \ \text{and}\ \ \ \ \bar\mu\to\infty\  
\end{equation}
with the finite dimensionless ratio
\begin{equation}\label{Eq4}
-{{\bar\eta}\over{{\bar\mu}^2}}\to\text{finite value}\  ,  
\end{equation}
where $\bar\mu\equiv M\mu$ is the dimensionless proper mass of the supported non-minimally coupled scalar field. 

\section{Description of the system}

We study the physical and mathematical properties of `cloudy' black-hole 
spacetimes which are composed of central spinning Kerr black holes that support spatially regular 
linearized bound-state configurations of massive scalar fields with a non-trivial (non-minimal) direct coupling to the 
Gauss-Bonnet invariant, 
\begin{equation}\label{Eq5}
{\cal G}\equiv R_{\mu\nu\rho\sigma}R^{\mu\nu\rho\sigma}-4R_{\mu\nu}R^{\mu\nu}+R^2\  ,
\end{equation}
of the curved black-hole spacetime. 

The action of the Einstein-Gauss-Bonnet-nonminimally-coupled-massive-scalar field theory is 
given by the integral expression \cite{GB1,GB2,Noteun}
\begin{equation}\label{Eq6}
S={1\over2}\int
d^4x\sqrt{-g}\Big[R-{1\over2}\nabla_{\alpha}\phi\nabla^{\alpha}\phi-{1\over2}\mu^2\phi^2+f(\phi){\cal
G}\Big]\  .
\end{equation}
Following \cite{GB1,GB2,Hersc1}, we shall assume that the non-minimal coupling
function of the massive scalar field is characterized by the leading order compact functional behavior
\begin{equation}\label{Eq7}
f(\phi)={1\over2}\eta\phi^2\  ,
\end{equation}
which guarantees that the familiar (scalarless) spinning Kerr black-hole spacetime [see Eq. (\ref{Eq9}) below] 
is a valid solution 
of the coupled Einstein-scalar field equations in the $\phi\to0$ limit \cite{GB1,GB2,Hersc1}. 
The strength of the direct non-minimal interaction between the supported massive scalar field configurations and the 
Gauss-Bonnet invariant (\ref{Eq5}) of the curved spacetime is controlled by the dimensionless 
physical parameter
\begin{equation}\label{Eq8}
\bar\eta\equiv {{\eta}\over{M^2}}\  .
\end{equation}
The weak-field coupling function (\ref{Eq7}) of the composed 
Einstein-Gauss-Bonnet-nonminimally-coupled-massive-scalar field theory (\ref{Eq6}) may be characterized by 
either positive \cite{ChunHer} or negative \cite{SotN} values of the non-minimal expansion parameter $\bar\eta$.

The supporting Kerr black-hole spacetime of mass $M$ and angular momentum $J\equiv Ma$ \cite{Noteaaa} 
is described, using the familiar Boyer-Lindquist coordinates, by the curved line element \cite{ThWe,Chan}
\begin{eqnarray}\label{Eq9}
ds^2=-{{\Delta}\over{\rho^2}}(dt-a\sin^2\theta
d\phi)^2+{{\rho^2}\over{\Delta}}dr^2+\rho^2
d\theta^2+{{\sin^2\theta}\over{\rho^2}}\big[a
dt-(r^2+a^2)d\phi\big]^2\  .
\end{eqnarray}
The spatially-dependent metric functions in (\ref{Eq9}) are given by the 
functional expressions
\begin{equation}\label{Eq10}
\Delta\equiv r^2-2Mr+a^2\ \ \ \ ; \ \ \ \ \rho^2\equiv r^2+a^2\cos^2\theta\  .
\end{equation}
The black-hole horizon radii,
\begin{equation}\label{Eq11}
r_{\pm}=M\pm(M^2-a^2)^{1/2}\  ,
\end{equation}
are determined by the radial roots of the metric function $\Delta(r)$. 

The supported massive scalar field configurations that we shall analyze in the present paper 
are assumed to be directly coupled [see the action (\ref{Eq6})] to the spatially-dependent 
Gauss-Bonnet invariant (\ref{Eq5}) which, in the spinning curved Kerr black-hole spacetime (\ref{Eq9}) can 
be expressed in the form \cite{SotN}
\begin{equation}\label{Eq12}
{\cal G}_{\text{Kerr}}(r,\theta)={{48M^2}\over{(r^2+a^2\cos^2\theta)^6}}
\cdot\big({r^6-15a^2r^4\cos^2\theta+15a^4r^2\cos^4\theta-a^6\cos^6\theta}\big)\  .
\end{equation}

The action (\ref{Eq6}), which characterizes the composed 
Einstein-Gauss-Bonnet-nonminimally-coupled-massive-scalar field theory, 
yields the generalized Klein-Gordon equation \cite{GB1,GB2,Hersc1,Donnw2,Notemassive}
\begin{equation}\label{Eq13}
\nabla^\nu\nabla_{\nu}\phi=\mu^2_{\text{eff}}\phi\  ,
\end{equation}
where the effective scalar-field-Gauss-Bonnet mass term, whose presence in (\ref{Eq13}) reflects 
the direct coupling between the scalar field and the Gauss-Bonnet curvature invariant (\ref{Eq5}), 
is given by the spatially-dependent ($\{r,\theta\}$)-dependent) functional expression
\begin{equation}\label{Eq14}
\mu^2_{\text{eff}}(r,\theta;M,a)=\mu^2-\eta\cdot{\cal G}_{\text{Kerr}}(r,\theta)\  .
\end{equation}

Intriguingly, taking cognizance of Eq. (\ref{Eq12}), one finds that, 
depending on the values of the physical parameters $\{M,a,\eta,\mu\}$ that characterize the composed black-hole-field system, 
the effective mass term (\ref{Eq14}) may become negative 
in the exterior region of the black-hole spacetime. In the next section we shall explicitly prove that 
this physically interesting property of the spatially-dependent effective mass term (\ref{Eq14}) allows the existence of composed 
black-hole-massive-scalar-field bound-state configurations in 
the non-trivial Einstein-Gauss-Bonnet-nonminimally-coupled-massive-scalar field theory (\ref{Eq6}). 

\section{Non-equatorial massive scalar rings supported by rapidly-spinning Kerr black holes}

In the present section we shall analyze the physical and mathematical properties of cloudy 
spinning-Kerr-black-hole-massive-scalar-field configurations that mark the 
{\it onset} of the spontaneous scalarization phenomenon in the composed 
Einstein-Gauss-Bonnet-nonminimally-coupled-massive-scalar field theory (\ref{Eq6}). 
In particular, we shall reveal the physically interesting fact that for negative values,
\begin{equation}\label{Eq15}
\bar\eta<0\  ,
\end{equation}
of the non-trivial dimensionless coupling parameter of the theory, rapidly-rotating 
black holes can support {\it non}-equatorial thin matter rings which 
are made of the non-minimally coupled massive scalar fields. 

To this end, we first point out that one deduces from Eq. (\ref{Eq12}) that, depending on the relative magnitudes of the physical 
parameters $\eta$ and $\mu$ that characterize the non-trivial field theory (\ref{Eq6}), 
the functional expression (\ref{Eq14}) for the effective mass term of the scalar field may become negative (thus 
representing an {\it attractive} black-hole-field binding potential well) in the 
exterior region of the spinning Kerr black-hole spacetime. 
This physically interesting property of the composed 
Einstein-Gauss-Bonnet-nonminimally-coupled-massive-scalar field theory allows the central 
black hole to support bound-state configurations of the non-minimally coupled 
massive scalar fields \cite{Donnw2,ChunHer,Hodca,Herkn}. 

In particular, in the dimensionless large-mass regime 
\begin{equation}\label{Eq16}
\bar\mu\equiv M\mu\gg1\
\end{equation}
of the composed black-hole-field system 
[or equivalently, in the regime $-\bar\eta\gg1$ of large negative values of the non-minimal coupling parameter, see Eq. (\ref{Eq22}) below], 
the {\it onset} of the spontaneous scalarization phenomenon in the non-trivial field theory (\ref{Eq6}) 
is marked by the critical functional behavior \cite{Hodca,Herkn,Hodjp}
\begin{equation}\label{Eq17}
\text{min}\{\mu^2_{\text{eff}}(r,\theta;M,a)\}\to 0^{-}\
\end{equation}
of the spatially-dependent effective mass term that characterizes the generalized Klein-Gordon 
equation (\ref{Eq13}) of the non-minimally coupled scalar field. 

Performing a functional analysis of the two-dimensional ($\{r,\theta\}$-dependent) expression (\ref{Eq12}) 
one finds that, for rapidly-rotating Kerr black holes in the dimensionless spin regime
\begin{equation}\label{Eq18}
{\bar a}\equiv\Big({{a}\over{M}}\Big)>\Big({{a}\over{M}}\Big)_{\text{crit}}=
\sqrt{{{7+\sqrt{7}\cos\Big[{1\over3}\arctan\big(3\sqrt{3}\big)\Big]-
\sqrt{21}\sin\Big[{1\over3}\arctan\big(3\sqrt{3}\big)\Big]}\over{12}}}
\  ,
\end{equation}
the Kerr Gauss-Bonnet curvature invariant is characterized by the extremum relation 
\begin{eqnarray}\label{Eq19}
{\text{min}\Big\{M^4\cdot{\cal G}_{\text{Kerr}}(r\in[r_+,\infty],\cos^2\theta;{\bar a})\Big\}}=
-{{57+28\sqrt{21}\cos\big[{1\over3}\arctan\big({{1}\over{3\sqrt{3}}}\big)\big]}\over{8(1+\sqrt{1-{\bar a}^2})^6}}\  ,
\end{eqnarray}
where the spin-dependent minimum (\ref{Eq19}) is characterized by the polar angular relation \cite{Notel1}
\begin{eqnarray}\label{Eq20}
(\cos^2\theta)_{\text{min}}= 
{{(1+\sqrt{1-{\bar a}^2})^2}\over{{\bar a}^2}}\cdot
\Big\{7-4\sqrt{7}\sin\Big[{1\over3}\arctan\Big({{1}\over{3\sqrt{3}}}\Big)\Big]-
4\sqrt{{{7}\over{3}}}\cos\Big[{1\over3}\arctan\Big({{1}\over{3\sqrt{3}}}\Big)\Big]\Big\}<1\
\
\end{eqnarray}
with 
\begin{equation}\label{Eq21}
r_{\text{min}}\to r_+(M,a)\  .
\end{equation}

It is interesting to note that the analytically derived expression (\ref{Eq20}) for $(\cos^2\theta)_{\text{min}}$ 
is a monotonically decreasing function of the black-hole dimensionless spin parameter ${\bar a}$ 
in its regime of validity (\ref{Eq18}). The data presented in Table \ref{Table1} clearly demonstrates 
the simple {\it monotonic} functional behavior of the spin-dependent angular 
function $(\cos^2\theta)_{\text{min}}$ [see Eq. (\ref{Eq20})] in the super-critical regime (\ref{Eq18}). 
In particular, the value of $|\theta_{\text{min}}({\bar a})|$ increases monotonically from $\theta_{\text{min}}({\bar a}={\bar a}_{\text{crit}})=0^\circ$ to $|\theta_{\text{min}}({\bar a}=1)|=61.21^\circ$.

\begin{table}[htbp]
\centering
\begin{tabular}{|c|c|c|c|c|c|c|c|c|}
\hline \ \ ${\bar a}$\ \ & \ $0.79$\ \ & \ $0.82$\ \ & \
$0.85$\ \ & \ $0.88$\ \ & \ $0.91$\ \ & \ $0.94$\ \ & \ $0.97$\ \ & \ $1.00$\ \ \\
\hline \ \ $[(\cos^2\theta)_{\text{min}}]({\bar a})$\ \ &\ \
$0.967$\ \ \ &\ \ $0.853$\ \ \ &\ \ $0.748$\ \ \ &\ \ $0.652$\ \ \ &\ \ $0.560$\ \ \ &\ \ 
$0.472$\ \ \ &\ \ $0.381$\ \ \ &\ \ $0.232$\ \ \\
\hline
\end{tabular}
\caption{Non-equatorial scalar rings supported by rapidly spinning Gauss-Bonnet black holes. 
We display, for various values of the black-hole dimensionless spin parameter ${\bar a}$ 
in the super-critical regime (\ref{Eq18}), the values of $[(\cos^2\theta)_{\text{min}}]({\bar a})$ as calculated from the analytically derived 
formula (\ref{Eq20}). The data presented demonstrates 
the simple {\it monotonic} functional behavior of the angular function (\ref{Eq20}). In particular, one 
finds that the value of $|\theta_{\text{min}}|$ increases monotonically from $\theta_{\text{min}}({\bar a}={\bar a}_{\text{crit}})=0^\circ$ to $|\theta_{\text{min}}({\bar a}=1)|=61.21^\circ$.} \label{Table1}
\end{table}

From Eqs. (\ref{Eq18}) and (\ref{Eq19}) one deduces that, in the dimensionless 
large-mass (or equivalently, large-coupling $-\bar\eta\gg1$) regime (\ref{Eq16}) with the critical 
relation \cite{Noteexc}
\begin{equation}\label{Eq22}
-{{57+28\sqrt{21}\cos\big[{1\over3}\arctan\big({{1}\over{3\sqrt{3}}}\big)\big]}\over{8(1+\sqrt{1-{\bar a}^2})^6}}
\cdot{{\bar\eta}\over{{\bar\mu}^2}}\to 1^{+}\  ,
\end{equation}
the effective mass term (\ref{Eq14}) that appears in the scalar Klein-Gordon equation (\ref{Eq13}) of the composed 
black-hole-field system becomes {\it negative} (thus representing an attractive black-hole-field potential well) 
in a pair of narrow {\it non}-equatorial rings which are characterized by the polar angular 
relation (\ref{Eq20}). 

This physically intriguing property of the composed 
Einstein-Gauss-Bonnet-nonminimally-coupled-massive-scalar field theory (\ref{Eq6}) 
implies that, in the large-mass (large-coupling) regime (\ref{Eq16}), 
rotating Kerr black holes in the dimensionless large-spin regime (\ref{Eq18}) 
can support pairs of non-equatorial thin matter rings [see Eqs. (\ref{Eq20}) and (\ref{Eq21})] 
which are made of massive scalar fields that are negatively 
coupled to the Gauss-Bonnet curvature invariant of the black-hole spacetime. 

In particular, the analytically derived functional expression (\ref{Eq22}) determines the spin-dependent 
critical existence-line $\bar\eta=\bar\eta(\bar\mu,{\bar a})$ that characterizes the onset of the spontaneous scalarization 
phenomenon in the composed 
Einstein-Gauss-Bonnet-nonminimally-coupled-massive-scalar field theory (\ref{Eq6}) in the dimensionless 
large-coupling-large-mass ($-\bar\eta\gg1,\bar\mu\gg1$) regime. 

\section{The classically allowed region of the non-equatorial near-critical massive scalar rings}

In the present section we shall use analytical techniques in order to 
determine the classically allowed spatial region of the supported non-minimally coupled 
massive scalar rings in the near-critical regime (\ref{Eq22}). 
To this end, it proves useful to use the dimensionless physical parameters $\{\epsilon,x,y\}$ which 
are defined by the following near-critical relations [see Eqs. (\ref{Eq20}), (\ref{Eq21}), and (\ref{Eq22})]
\begin{equation}\label{Eq23}
{{\bar\eta}\over{{\bar\mu}^2}}=\Big({{\bar\eta}\over{{\bar\mu}^2}}\Big)_{\text{crit}}\cdot(1+\epsilon)\ \ \ ; \ \ \ 
\Big({{\bar\eta}\over{{\bar\mu}^2}}\Big)_{\text{crit}}=
-{{8(1+\sqrt{1-{\bar a}^2})^6}\over{57+28\sqrt{21}\cos\big[{1\over3}\arctan\big({{1}\over{3\sqrt{3}}}\big)\big]}}
\ \ \ ; \ \ \ 0\leq\epsilon\ll1\  ,
\end{equation}
\begin{equation}\label{Eq24}
\cos^2\theta=(\cos^2\theta)_{\text{min}}\cdot(1+x)\ \ \ ; \ \ \ x\ll1\  ,
\end{equation}
and 
\begin{equation}\label{Eq25}
r=r_+\cdot(1+y)\ \ \ ; \ \ \ 0\leq y\ll1\  .
\end{equation}

Substituting the functional relations (\ref{Eq23}), (\ref{Eq24}), and (\ref{Eq25}) into the expression (\ref{Eq14}) for 
the spatially-dependent [see (\ref{Eq12})] effective mass term that characterizes the composed 
spinning-Kerr-black-hole-nonminimally-coupled-massive-scalar-field system, one finds that the black-hole-field 
effective interaction potential is negative (thus representing an {\it attarctive} black-hole-field binding potential) 
in a pair of narrow non-equatorial rings which are characterized by the dimensionless functional 
relation
\begin{equation}\label{Eq26}
\gamma x^2+6y\leq\epsilon\  ,
\end{equation}
where 
\begin{equation}\label{Eq27}
\gamma\equiv -{{6\beta^2(\beta^3-27\beta^2+63\beta-21)}\over{(1+\beta)^2(\beta^3-15\beta^2+15\beta-1)}}>0\
\end{equation}
with 
\begin{eqnarray}\label{Eq28}
\beta\equiv
7-4\sqrt{7}\sin\Big[{1\over3}\arctan\Big({{1}\over{3\sqrt{3}}}\Big)\Big]-
4\sqrt{{{7}\over{3}}}\cos\Big[{1\over3}\arctan\Big({{1}\over{3\sqrt{3}}}\Big)\Big]
\  .
\end{eqnarray}
The functional relation (\ref{Eq26}) with the definitions (\ref{Eq24}) and (\ref{Eq25}) determine, in the 
near-critical regime (\ref{Eq23}), the classically allowed spatial region for the spontaneous scalarization phenomenon of negatively-coupled massive scalar 
fields in rapidly-spinning Kerr black-hole spacetimes [see (\ref{Eq18})]. 

Interestingly, from Eqs. (\ref{Eq24}), (\ref{Eq25}), and (\ref{Eq26}) one deduces that, for a given value of the dimensionless 
near-critical parameter $\epsilon$ [with $\epsilon\ll1$, see Eq. (\ref{Eq23})], 
the maximum effective angular width of the supported non-equatorial 
scalar rings is given by the functional relation \cite{Noteang}
\begin{equation}\label{Eq29}
[\Delta(\cos^2\theta)]_{\text{ring}}={{(1+\sqrt{1-{\bar a}^2})^2}\over{{\bar a}^2}}\cdot
\sqrt{-{{2(1+\beta)^2(\beta^3-15\beta^2+15\beta-1)}\over{3(\beta^3-27\beta^2+63\beta-21)}}}\cdot\sqrt{\epsilon}
\ \ \ \ ; \ \ \ \ \epsilon\ll1\  ,
\end{equation}
and the maximum effective radial width of the supported scalar rings is 
given by the relation \cite{Noterad}
\begin{equation}\label{Eq30}
(\Delta r)_{\text{ring}}={1\over6}\cdot r_+\cdot\epsilon\ \ \ \ ; \ \ \ \ \epsilon\ll1\  .
\end{equation}

\section{Summary and Discussion}

The recently published physically interesting studies \cite{Sot5,Sot1,GB1,GB2} (see also \cite{ChunHer,SotN,Hodsg1,Hodsg2,Done,Herrecnum,BeCo}) 
of the composed Einstein-Gauss-Bonnet-scalar field theory have revealed that 
black holes with spatially regular horizons can support hairy matter configurations which are 
made of scalar fields with a non-trivial (non-minimal) coupling to the spatially dependent 
Gauss-Bonnet invariant ${\cal G}$ of the curved spacetime. 

Intriguingly, it has been demonstrated \cite{ChunHer,SotN} that, 
for spinning black holes in Einstein-Gauss-Bonnet-scalar field theories, the sharp 
boundary between the familiar (scalarless) Kerr black-hole spacetime and spontaneously scalarized hairy black-hole 
spacetimes is determined by the presence of cloudy black-hole-field configurations, Kerr black holes that support linearized 
bound-state configurations of the spatially regular non-minimally coupled scalar fields. 
In particular, the composed black-hole-field cloudy configurations determine 
a universal [that is, independent of the non-linear functional behavior of the scalar coupling function 
$f(\phi)$, see (\ref{Eq6})] critical existence-line $\bar\eta_{\text{crit}}=\bar\eta_{\text{crit}}({\bar a})$ that describes the non-trivial 
spin-dependent functional behavior of the non-minimal coupling parameter that characterizes the composed 
Einstein-Gauss-Bonnet-scalar field theory. 

In the present paper we have used {\it analytical} techniques in order to explore 
the onset of the negative-coupling ($\bar\eta<0$) spontaneous scalarization phenomenon of 
spinning Kerr black-hole spacetimes in the large-coupling-large-mass 
regime
\begin{equation}\label{Eq31}
\{-\bar\eta,\bar\mu\}\to \infty\ \ \ \ \text{with}\ \ \ \ -{{\bar\eta}\over{{\bar\mu}^2}}\to\text{finite value}\  .  
\end{equation}
The main results derived in this paper and their physical implications are as follows: 

(1) We have revealed the physically interesting fact that in 
the Einstein-Gauss-Bonnet-nonminimally-coupled-massive-scalar field theory (\ref{Eq6}), 
rapidly-rotating Kerr black holes in the large-spin regime 
\begin{equation}\label{Eq32}
{\bar a}>{\bar a}_{\text{crit}}\
\end{equation}
with [see Eq. (\ref{Eq18})]
\begin{equation}\label{Eq33}
{\bar a}_{\text{crit}}=
\sqrt{{{7+\sqrt{7}\cos\Big[{1\over3}\arctan\big(3\sqrt{3}\big)\Big]-
\sqrt{21}\sin\Big[{1\over3}\arctan\big(3\sqrt{3}\big)\Big]}\over{12}}}
\simeq0.78\
\end{equation}
can support a pair of {\it non}-equatorial thin matter rings.

(2) It has been shown that 
the supported non-minimally coupled massive scalar rings are characterized by the analytically derived dimensionless 
angular relation [see Eqs. (\ref{Eq11}) and (\ref{Eq22})]
\begin{eqnarray}\label{Eq34}
\big({{a}\over{r_+}}\big)^2\cdot(\cos^2\theta)_{\text{ring}}= 
7-4\sqrt{7}\sin\Big[{1\over3}\arctan\Big({{1}\over{3\sqrt{3}}}\Big)\Big]-
4\sqrt{{{7}\over{3}}}\cos\Big[{1\over3}\arctan\Big({{1}\over{3\sqrt{3}}}\Big)\Big]
\  .
\end{eqnarray}

(3) We have proved that the negatively-coupled non-equatorial scalar rings are characterized by the 
large-mass-large-coupling 
critical ratio [see Eqs. (\ref{Eq11}) and (\ref{Eq22})] 
\begin{equation}\label{Eq35}
\Big(-{{\bar\eta}\over{{\bar\mu}^2}}\Big)_{\text{crit}}\to
\Bigg\{{{8}\over{57+28\sqrt{21}
\cos\big[{1\over3}\arctan\big({{1}\over{3\sqrt{3}}}\big)\big]}}\cdot\Big({{r_+}\over{M}}\Big)^6\Bigg\}^+\  .
\end{equation}
It is worth emphasizing again that the physical significance of the 
analytically derived spin-dependent dimensionless functional relation (\ref{Eq35}), which characterizes 
the composed Kerr-linearized-massive-scalar-field cloudy configurations, stems from the fact that this 
critical existence-line determines the sharp boundary between bald Kerr black holes and 
spontaneously scalarized hairy spinning black holes in the 
Einstein-Gauss-Bonnet-nonminimally-coupled-massive-scalar field theory (\ref{Eq6}).

(4) It has been shown that, in the near-critical regime 
\begin{equation}\label{Eq36}
{{\bar\eta}\over{{\bar\mu}^2}}=\Big({{\bar\eta}\over{{\bar\mu}^2}}\Big)_{\text{crit}}\cdot(1+\epsilon)
\ \ \ ; \ \ \ 0\leq\epsilon\ll1\
\end{equation}
of the composed Einstein-Gauss-Bonnet-nonminimally-coupled-massive-scalar field theory (\ref{Eq6}), 
the maximum effective spatial widths of the 
supported non-equatorial scalar rings are characterized by the small-$\epsilon$ dimensionless 
functional relations [see Eqs. (\ref{Eq29}) and (\ref{Eq30})]
\begin{equation}\label{Eq37}
{{{\bar a}^2}\over{{(1+\sqrt{1-{\bar a}^2})^2}}}\cdot[\Delta(\cos^2\theta)]_{\text{ring}}=
\sqrt{-{{2(1+\beta)^2(\beta^3-15\beta^2+15\beta-1)}\over{3(\beta^3-27\beta^2+63\beta-21)}}}\cdot\sqrt{\epsilon}
\ \ \ \ ; \ \ \ \ {{(\Delta r)_{\text{ring}}}\over{r_+}}={1\over6}\cdot\epsilon\  .
\end{equation}

\bigskip
\noindent
{\bf ACKNOWLEDGMENTS}
\bigskip

This research is supported by the Carmel Science Foundation. I would
like to thank Yael Oren, Arbel M. Ongo, Ayelet B. Lata, and Alona B.
Tea for helpful discussions.



\begin{thebibliography}{99}
\bibitem{Notechar} See \cite{Hersc1,Hersc2,Hodsc1,Hodsc2,Moh,Hodnrx,HodKN} for the physically related case 
of charged black hole in composed Einstein-Maxwell-scalar field theories that can support spatially regular bound-state 
configurations of static scalar fields which are directly (non-minimally) coupled to the spatially-dependent 
Maxwell electromagnetic invariant.
\bibitem{Notelll} It is worth noting that, as explicitly proved in \cite{Hodca}, the critical 
black-hole spin (\ref{Eq1}) corresponds to the regime $l\gg1$. 
In particular, one finds \cite{Hodca} that for smaller $l$ values the critical rotation parameter 
that allows for the existence of negatively coupled scalar clouds is larger than the critical value (\ref{Eq1}).
\bibitem{Noteun} We use natural units in which $G=c=\hbar=1$. Note that in these units the mass $\mu$ of the 
non-minimally coupled scalar field has the dimensions of length$^{-1}$. The black-hole-scalar-field composed 
mass parameter $\bar\mu\equiv M\mu$ is dimensionless.
\bibitem{Noteaaa} We shall assume, without loss of generality, the relation ${\bar a}\equiv J/M^2>0$ for the angular momentum 
parameter of the supporting Kerr black hole.
\bibitem{Notemassive} It is worth emphasizing that the fact that the non-minimally coupled scalar field 
has a finite mass allows one to obtain non-equatorial scalar rings whose widths can be made arbitrarily thin. 
In particular, we shall explicitly prove below that, by adjusting properly the mass term 
of the non-minimally coupled scalar field [see Eq. (\ref{Eq22}) below], the classically allowed region 
of the composed black-hole-field effective potential term (\ref{Eq14}) 
[the region in which the effective mass term (\ref{Eq14}) is negative and therefore represents a binding potential well] 
can be made arbitrarily narrow, thus allowing for the existence of arbitrarily thin non-equatorial 
scalar rings which are characterized by the property (\ref{Eq17}).
\bibitem{Notel1} Note that the inequality in (\ref{Eq20}) is valid in the super-critical regime (\ref{Eq18}).
\bibitem{Noteexc} Note that the critical dimensionless ratio (\ref{Eq22}) 
corresponds to the fundamental composed black-hole-field configuration. 
The composed black-hole-field system also has excited bound states which are characterized 
by the near-critical relation (\ref{Eq23}) with $\epsilon>0$.
\bibitem{Noteang} Note that the angular width is defined as the angular interval in which the effective 
mass term (\ref{Eq14}) in the scalar equation (\ref{Eq13}) 
is negative and therefore represents an attractive binding potential well of 
the composed black-hole-field system.
\bibitem{Noterad} Note that the radial width is defined as the radial interval in which the effective 
mass term (\ref{Eq14}) is negative and therefore represents an attractive black-hole-field binding potential well.
\end{thebibliography}
\end{document}